# Data-driven modeling of systemic delay propagation under severe meteorological conditions

Pablo Fleurquin[1,2], José J. Ramasco[1] and Victor M. Eguiluz[1]

[1]Instituto de Física Interdisciplinar y Sistemas Complejos IFISC (CSIC-UIB), Campus UIB, 07122 Palma de Mallorca, Spain and
[2]Innaxis Foundation & Research Institute, José Ortega y Gasset 20, 28006 Madrid, Spain.

**The upsetting consequences of weather conditions are well known to any person involved in air transportation. Still the quantification of how these disturbances affect delay propagation and the effectiveness of managers and pilots interventions to prevent possible large-scale system failures needs further attention. In this work, we employ an agent-based data-driven model developed using real flight performance registers for the entire US airport network and focus on the events occurring on October 27 2010 in the United States. A major storm complex that was later called the 2010 Superstorm took place that day. Our model correctly reproduces the evolution of the delay-spreading dynamics. By considering different intervention measures, we can even improve the model predictions getting closer to the real delay data. Our model can thus be of help to managers as a tool to assess different intervention measures in order to diminish the impact of disruptive conditions in the air transport system.**

## I. INTRODUCTION

Flight delays drained $40.7 billion from the U.S. economy in 2007 [1]. Airlines economic results sustained expenses equivalent to approximately one half of the previous amount due to an increase in the operating costs. Likewise, similar costs can be expected in Europe, with billions wasted in system inefficiencies [2,3]. The impact of flight delays is not only economic but also environmental due to additional $CO_2$ emissions to recover delays [4]. Understanding how delays generate and propagate across the transport network involves considering a large number of variables and elements. The basic elements internal to the system include, of course, the flights but also passengers, crews, airport operations, etc. Additionally, other external factors can affect flight performance as, for example, weather, labor regulations and strikes or security threats. The intricacy of all these elements and of the interactions between them clearly qualifies ATM as an area to be considered under the light of Complex Systems theory. Complexity is characterized by the emergence of new phenomena as a result of the interactions between the elements of a system. Several tools have been introduced to study and quantify the properties of such emergent behaviors. In this case, we use these tools and focus on flight delay propagation. Given the natural networked structure of the system, we take a holistic approach and tackle this problem from a network-wide perspective. To do so, we define metrics that measure the macro-scale behavior of the delay dynamics. Clusters of interconnected delayed airports are constructed in order to assess the level of system congestion and the importance of network connectivity in the unfolding of the delay spreading mechanism.

In a recent work [5], we have developed a realistic agent-based model able to reproduce the delay distribution patterns observed in the real performance data for the US. In that work, we assessed the influence that internal factors (aircraft rotation, airport congestion, passenger connections and crew rotation) have on the performance of the system. Based on these findings we extend the use of our model to understand the system response to the introduction of large-scale disruptions and to provide a tool to assess strategies to handle these disturbances. Such system disruptions are commonly caused by adverse weather, ranging from reduced ceiling and visibility to convective weather. Therefore, we selected for simulation and comparison purposes October 27[th] 2010, a day for which we count with real performance data and that turned to be the worst day in 2010 according to the average flight delay. The origin of this high congestion levels was a severe weather phenomena distributed across the country [6,7]. Indeed, the meteorological disturbances of this day were later known in the media as part of the 2010 Superstorm.

The remainder of the paper is organized as follows. A background review of the literature on delay propagation is provided in Section II. Section III is devoted to describe the database. The model elements and structure are discussed in Section IV. The results are presented in Section V. And, finally, Section VI contains the paper conclusions.

## II. BACKGROUND

There are several definitions of flight delays. For instance, according to the FAA a flight can be considered as delayed if the operation takes place 15 minutes after scheduled [8]. In our work, we follow the definition given in [9,10] and define delay as the time difference between real and scheduled operations (arrival or departure). This definition is more flexible and does not filter out small delays that can form part of a general state of congestion.

Flight delay propagation has received a lot of attention from the ATM community during the last decade. A significant effort has been invested in identifying the causes for initial or primary delays [10,11]. Among the sources of primary delay, some of the most devastating are related to weather perturbations as has been shown in [12-14]. These primary delays can in turn trigger a cascade of secondary delays as was noted in [9] by the introduction of a ripple effect. The primary delays affecting early flights could propagate downstream and induce delays to several flights later in the day, the so-called reactionary delay [15-17]. This cascade-like effect is produced and magnified by the network connectivity through the aircraft rotation and the crew and passenger connections between flights. Our previous findings suggest that crew and passenger connectivity is the most damaging internal factor spawning system-wide effects [5].

The linkage between flights is an essential feature of the networked structure of the air traffic system. Therefore the propagation dynamics cannot be understood without referring to the underlying complex structure. The use of network theory to characterize air transportation describes the system as a graph formed with vertices representing commercial airports and edges direct flights between them. The initial works [18,19] found a high heterogeneity in the traffic sustained by each edge and the number of connections per airport. Furthermore, in [18] a correlation between the network topology (node degree) and the number of travelers has been observed. In addition, it was found that the air traffic network is structured in communities of airports that reflect geographical areas with high internal traffic [20]. The dynamics of the networks has been also studied in [21]. All these aspects affect the way in which delay propagates across the network.

Because of the inherent complexity of the mechanisms that produce and boost delay spreading, different modeling techniques were proposed. A first line of research focused on simulating the air traffic system as a network of queues without considering information on aircraft schedules [22]. A second line of research was devoted to analytical approximations for modeling the airport runway operations as a dynamic queuing system with varying demand and service rate [23]. Another analytical queuing model was used in [24]. In this work, airports were modeled as dynamic queues and implemented in a network. The authors ran the model in a network of 34 airports with a specific algorithm that accounts for downstream propagation of delays. An additional body of work uses statistical tools to predict the delay patterns observed in the data. Such techniques could be classified into traditional linear regression models [25], artificial neural networks [26] and Bayesian networks [27].

Some of the mentioned models have been limited to single-airport or just a few major airports analysis with different level of detail, while others performed an aggregated analysis of the whole system. By considering an agent-based framework [28] we can give insights, in a cost-effective way, of how micro-level interactions give place to emergent behavior from a network-wide perspective. In this sense, we have analyzed the system response to the introduction of primary delays in the first flight leg of an aircraft itinerary. By going one step further, in this work, we want to understand the system behavior to disrupting events that may compromise the system stability. Previous attempts to understand the stability of the air traffic network were based on queuing [29] and percolation [30] theory. Here we simulate the system performance under weather-disrupting inputs modeled as a shortfall on terminal capacity. Our viewpoint falls in line with earlier research on weather-related delays associated to capacity constraints [12,13].

### III. DATA SOURCES

The data sources used for the analysis can be categorized into three distinct groups: delay data, time zone conversion data and airport connectivity data.

#### A. Delay data

Delay data was obtained from the Bureau of Transport Statistics [31]. Specifically the information was drawn from the Airline On-Time Performance Data, which is composed with flight data provided by air carriers that exceeds one percent of the annual national revenue for domestic scheduled service. Added together this represents 18 carriers that combined sum up 6,450,129 scheduled domestic flights operated from 305 commercial airports. The total number of flights in the US for 2010 is 8,687,800 [32]. Consequently, the database includes information accounting for the 74% of the total. Among the flight information provided, we use for modeling purposes the aircraft and airline identification code (tail number and airline id, respectively), date of flight, real and scheduled departure (arrival) times, origin and destination, and whether the flight was canceled or diverted. In order to construct the US airport network and replicate the aircraft rotations for each day we combined the tail number code with the spatiotemporal localization provided by the dataset, excluding canceled and diverted flights. It is worth noting that the reconstructed aircraft itineraries are based on real events, which may differ from the original planned schedule of the airlines at the beginning of the day. Regarding this point we cannot trace back the original flight plan but one can thus expect these modifications to be of a small magnitude. This conclusion is sustained by the fact that canceled and diverted flights represent -on average-, respectively 1.75% and 0.20% of the total flights in the database. However, on modeling weather-impacted days, as it is the case, this assumption should be taken carefully and expect that this type of interventions on the network may modify the delay dynamics.

#### B. Time zone convertion

Since the operation time registers in the On-Time Performance Data are in local time, there is a need to unify criteria for the correct timing of the daily aircraft sequences. Because the United States spans through several time zones, we used the Olson or tz database [33] to ensure a correct conversion from the respective local times in the database to the East Coast local time (EST in winter and EDT in summer). We chose this time zone to follow the natural daylight time flow in the United States.

#### C. Airport connectivity data

Although the delay data allow us to reconstruct the aircraft itineraries, we lack information regarding crew and passenger connections. To address this issue, we can at least estimate the airport heterogeneities concerning flight connectivity. By pulling from the BTS data repository [31] the T100 Domestic Market and the DB1B Ticket information datasets we estimate the annual fraction of connecting passengers in each airport. The DB1B is a 10% sample of the number of passengers that started their flight plan in a given airport ($Passengers_{DB1B}$).

On the other hand, the T100 dataset provides information on the total number of passengers that start their journey from an airport regardless of their origin airport (Passengers$_{T100}$). Hence, we estimate the annual fraction of connecting passengers for each airport as:

$$\text{airport connectivity factor} = \frac{\text{Passengers}_{T100} - 10 \cdot \text{Passengers}_{DB1B}}{\text{Passengers}_{T100}}. \quad (1)$$

It is important to note that the approximation is only related to passenger connections and it does not take into account the crew linkage in subsequent flight legs. Even though our model accounts for flight, not only passenger, connectivity we assume that there is an intrinsic relationship between both.

IV. MODEL DESCRIPTION

We developed a data-driven model by using as inputs real records obtained from the different databases. By these means, the fundamental modeling structures, namely: the daily schedule and the airport network, explicitly incorporate the complex nature of the air traffic system into the modeling framework. In this sense, the daily schedule includes the planned slack time that should help mitigate the delay spreading mechanism. The airport network built from the data is composed of 305 (this value slightly varies from day to day) commercial airports and 2,318 edges. Also the heterogeneities present in this network matches the characteristics of complex networks reported in the literature. We avoid making arbitrary assumptions for the initial model input. The values used are the real primary delays of the first flight legs of each aircraft itinerary. The overall empirical approach, combined with an agent-based framework, enables the development of a model with predictive power that captures the real features of the delay dynamics, while maintaining a simple structure and few parameters.

The temporal resolution of the model is at the level of minutes, so we run the simulation using one minute as the basic time step. This time unit is the finest resolution available in the data. Each aircraft (agent) is tracked at this temporal resolution and each simulation proceeds until the planning of a selected day is fulfilled. Each aircraft is unique and could be track by their tail number code. At the "micro-level" three main sub-processes rule the agents' responses to each other and their environment: aircraft rotation, flight connectivity and airport congestion. The aircraft rotation is the basic ingredient of the schedule so it cannot be turned off. On the other hand, the remaining sub-processes were defined to explore the contribution of each one of them to the delay dynamics. Therefore, we can switch them on/off or moderate their importance independently. Depending on the day, approximately 4,000 airplanes participate in one simulation run completing an itinerary that, in most cases, is composed of more than one flight leg. This itinerary subdivision is the basic schedule unit. In other words, the minimum amount of information needed to move an airplane from an origin to a destination airport, according to a pre-established schedule. Naturally, an aircraft rotation is finished if all the constituent flight legs have been fulfilled in a sequential order. We considered a flight as not finished whenever the aircraft is in the gate-to-gate or block-to-block phase. In particular, in this phase it is not possible to absorb the delay, thus departure and arrival delays, for each flight leg, are the same. Opposed to the block-to-block time period, we consider the turn-around phase as the time that the aircraft remains parked at the gate [34]. Is in this phase where the delay reduction may occur if there is enough slack time without capacity shortfalls. This turn-around phase helps thus to maintain the aircraft sequence stability and the airport operational performance [35]. Another characteristic related to the aircraft rotation sub-process is that each aircraft, when arrived, must complete a minimum service time $T_s$ related to ground operations. This process includes de-boarding, fueling, luggage handling, cleaning, catering and boarding. For this sub-process considering an aircraft ($p_{ij}$) that flown from an origin $i$ to a destination airport $j$ the actual departure time of the next flight leg in j is given by:

$$T_{act.d}^j(p_{ij}) = \max[T_{sch.d}^j(p_{ij}); T_{act.a}^j(p_{ij}) + T_s]. \quad (2)$$

The sub-indices $act.d$, $act.a$ and $sch.d$ correspond respectively to Actual Departure, Actual Arrival and Schedule Departure time.

A secondary feature of the agents in the simulation is the air carrier identified by the airline code. This is a collection of aircrafts that interact through the flight connectivity sub-process. Only aircrafts belonging to the same carrier have a probability of connection proportional, with a factor α, to the connectivity levels of each airport. Specifically, for each scheduled departing flight in a particular airport a connection is randomly chosen with probability $\alpha \times$ (*airport connectivity factor*). We consider potential connections to flights of the same air carrier that, in addition, have a scheduled arrival time within a $\Delta T = 3$ hours time window prior to the departing flight under analysis. Therefore, this sub-process is responsible for the introduction of stochasticity into the model. Considering flight connectivity, an aircraft is allowed to fly if and if only its connections have already arrived to the airport. In this case the Actual Departure time of the next flight leg is given by:

$$T_{act.d}^j(p_{ij}) = \max\left[T_{sch.d}^j(p_{ij}); T_{act.a}^j(p_{ij}) + T_s; \max\left(T_{act.a}^j(p_{i'j})\right)\right], \forall\, i' \neq i. \quad (3)$$

The index $i'$ corresponds to the connections that the flight has to wait in order to depart.

The above interactions between airplanes take place at an airport. This is an intermediate-level object with unique features: location, airport connectivity factor and planned capacity. The latter characteristic is introduced in the model by computing the scheduled airport arrival rate for each hour (SAAR). To modulate this capacity, we include another proportionality factor β. Due to delays the airport demand profile may change and the real airport arrival rate may surpass the scheduled rate. In this case, a queue begins to form that may congest the airport generating even further delays. The implemented queuing protocol is "First in- First Served" a common queue operation procedure. When aircraft rotation and airport congestion are present the dynamics are ruled by:

$$T_{act.d}^j(p_{ij}) = \max[T_{sch.d}^j(p_{ij}); T_q^j + T_{act.a}^j(p_{ij}) + T_s]. \quad (4)$$

Where $T_q^j$ means the time spent by the aircraft in the queue waiting to be served. Finally, the full model dynamics is govern by a combination of the three sub-processes:

$$T_{act.d}^j(p_{ij}) = \max\left[T_{sch.d}^j(p_{ij}); T_q^j + T_{act.a}^j(p_{ij}) + T_s; \max\left(T_{act.a}^j(p_{i'j})\right)\right], \forall\, i' \neq i. \quad (5)$$

From our previous work we found that airport congestion by itself is not capable of spreading the delays and generating system-wide effects as it is the case of flight connectivity, but it plays an important role as a source of new "primary" delays.

### A. Input variables

The simulation begins at 4am Eastern Time when there is almost no traffic activity throughout the network. Also this time ensures that most aircraft itineraries are sorted correctly. The initial delays are introduced in three different ways. The first one takes as input variables the primary delay of the first flight of an aircraft sequence, replicating exactly the location and the timing of the first delay. The second procedure, reshuffles the primary delay among the first flight legs of each aircraft rotation, so when and where will vary. Finally, the model can track similar initial flights from one day and include their initial delays into the first flight legs of another day.

### B. Output variables, clusters of congested airports

As a way of characterizing the unfolding of the delay spreading in the network, we defined a high-level entity for the full system. First, we calculate a network per day form with the airports as nodes and the direct flights between them as connections. Then, we measure dynamical clusters that are constituted by congested airports whose average departure delay per flight exceeds a certain threshold and are connected in the network. This threshold amounts 29 minutes and corresponds to the average departure delay of all delayed flights during 2010. Note that congested clusters are related to spatio-temporal correlations but not to a cause-effect relation. By these means the model is able to compute, for instance the evolution of the largest cluster of the day or the number of clusters for each hour in order to compare it with the empirical results.

### C. Summary of model parameters

The model has two main free parameters: β, controlling the airport service capacity rate, and α, accounting for the flight connectivity. In the model simulations we set β to 1 assuming nominal capacity, in other words, the same airport capacity as originally scheduled. We then run the simulations and fit α to obtain the maximum cluster size as observed in the data. We also test the effect of varying other parameters and found that the model outcome was insensitive to these variations. For this reason, the other value parameters were selected because they match values reported in the literature of flight delays: $T_s = 30$ min and $\Delta T = 180$ min.

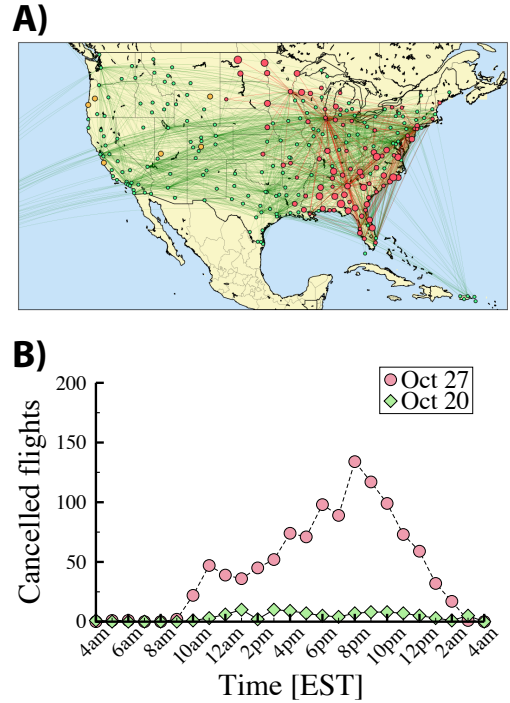

Figure 1. A) Map of the US airport network for October 27 showing the daily largest congested cluster in red. The airport color codes are: green, airport not congested; orange, congested airport not belonging to the largest cluster; red, congested airport belonging to the largest cluster. B) Number of canceled flights per hour.

## V. RESULTS

We compare next the model predictions with the real evolution of the delay events on October 27 2010. This day embodies the concept of external disturbances affecting a large part of the air transport network. The disturbances were the result of severe weather conditions generated by a low-pressure system that started during the early morning hours of the 26th in the Southern Plains and moved North, producing a significant pressure gradient that caused strong wind gusts [36]. The massive storm complex continued throughout the 27th and dissipated on the 28th impacting some airports of the National Aviation System. According to the news reports, at least Hartsfield-Jackson airport in Atlanta (ATL), and the three main airports of the New York-New Jersey area, John F. Kennedy airport (JFK), La Guardia airport (LGA) and Newark airport (EWR), experienced large delays because of inclement weather [7,37].

TABLE I. VARIATIONS IN THE MODEL

| Variant of the model | Characteristics (changes implemented) |
|---|---|
| Basic model | The airport capacity β equals 1 (nominal capacity) for every airport in the network and the flight connectivity factor α is 0.26. These values remain constant througout the day. |
| Baseline model (basic model + perturbation) | The airport capacity β equals 1 except for ATL, JFK, LGA and EWR that the value is set to 0 between 8 a.m. to 10 a.m. (Eastern Time). Flight connectivity α is 0.26. |
| Variant 1 (baseline model + intervention in the network) | Same β conditions as the Baseline model. Flight connectivity α equals 0.26 except for the time period between 7 p.m. and 9 p.m. (Eastern Time) that drops to 0. |
| Variant 2 (baseline model + intervention in the network) | Same β conditions as the Baseline model. Flight connectivity α equals 0.26 except for the time period between 7 p.m. and 9 p.m. (Eastern Time) that drops to 0.13. |
| Variant 3 (baseline model + intervention in the network) | Same β conditions as the Baseline model. Flight connectivity α equals 0.26 except for the time period between 6 p.m. and 10 p.m. (Eastern Time) that drops to 0. |
| Variant 4 (baseline model modified) | Flight connectivity α is 0.26. The airport capacity β equals 1 for every airport in the network. In this case the perturbation is included by issuing Ground Stops to flights whose destination is ATL, JFK, LGA and EWR between 8 a.m. to 10 a.m. (Eastern Time). |

Focusing on the On-Time Performance data, this day presented the largest average flight departure and arrival delay of the whole year. These values are, respectively, 54 and 53 minutes. Moreover, the largest congested cluster of the day amounts to 88 airports, the third worst performance of 2010 after March 12 (97 airports) and December 12 (103 airports). A map of the congested cluster of airports for October the 27th is shown on Fig. 1-A. The congested airports are plotted in orange if they do not belong to the largest cluster and in red if they do. The size of the symbols of the airports in the largest congested cluster is proportional to the average delay. In the map, we can observe how the network congestion affected a vast area that spread from Central to Eastern U.S matching the area where the windstorm developed. In addition, the average departure delay for the first legs of the day for the flight rotations equals 46 minutes, which ranks among the top 10 days with worst initial conditions. Remember that these primary delays are the initial conditions introduced as external inputs for the simulations of our model.

The map in Fig. 1-A graphically shows how weather perturbations can produce system-wide effects in the air transportation network. The managers and pilots of the different airlines and airports, of course, reacted to the problems generated by the weather disturbances. Typically, these reactions included flight delay but also cancellations and flight diversion to airports different from those of destination. These two latter factors introduce changes in the planned schedule that are important to analyze. We depict in Fig. 1-B the number of canceled flights per hour for October 27 and October 20. October 20 was a low congested day with only 2 airports being part of the daily largest congested cluster and also showed a low average flight delay equal to 24 minutes (less than half of the value for the 27th). From Fig. 1-B we can conclude that the rise of network congestion, if compared with a low congested regime, induced an important rate of flight cancellation.

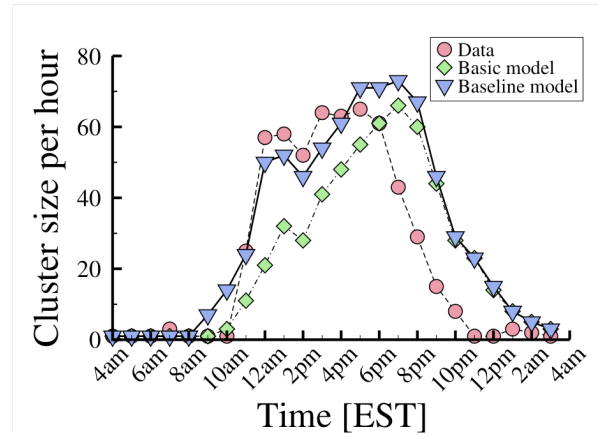

Figure 2. Evolution of the largest cluster size per hour. Comparison between reality (data), model without perturbation (basic) and model including perturbation (baseline).

The intraday evolution of the size of the congested cluster per hour is displayed in Figure 2. The cluster size shows an initial steep growing phase that starts at early morning hours and continues up to 5pm followed by a declining phase from 5pm onwards. We first used a simple model scenario with a fitted α value of 0.26 and introducing the initial conditions as given in the data to the schedule of October 27 (green dots Fig. 2). In this basic scenario, we consider every airport working at a nominal capacity (β = 1) throughout the day. Although the maximum cluster size is reproduced well, the simulation results show a slower increase in the growing phase that does not match the real evolution of the cluster size. In fact, it seems that this particular day morning hours are crucial to understand the development of the congested regime. In addition, the declining phase decays earlier in the empirical data. This difference regarding the dynamics could be due to the fact that the basic model does not take into account external perturbations to the system. As stated before, the severe weather conditions enhanced the delay spreading by affecting some airports capacity. As mentioned in [5] events of this kind can be included in the model by modifying the capacity parameter β. Although a change in β is not able to generate a network-wide spreading of the delays, it could be in fact a source of new "primary" delays that later on will propagate throughout the system. The new delay results from an increase in the length of the waiting queue mechanism. In this sense, we mimic the capacity shortfall of 4 airports (ATL, JFK, LGA and EWR), mentioned in the news report, by setting to 0 the capacity parameter in the morning hours (from 8am to 10am local time). This modification introduced to the basic model is shown in Fig. 2 as the baseline model. On the other hand, the connectivity factor remains constant. As shown in Fig. 2 our assumptions regarding the modeling of weather impacts through the β coefficient are confirmed, remarkably improving the model results in the growing phase.

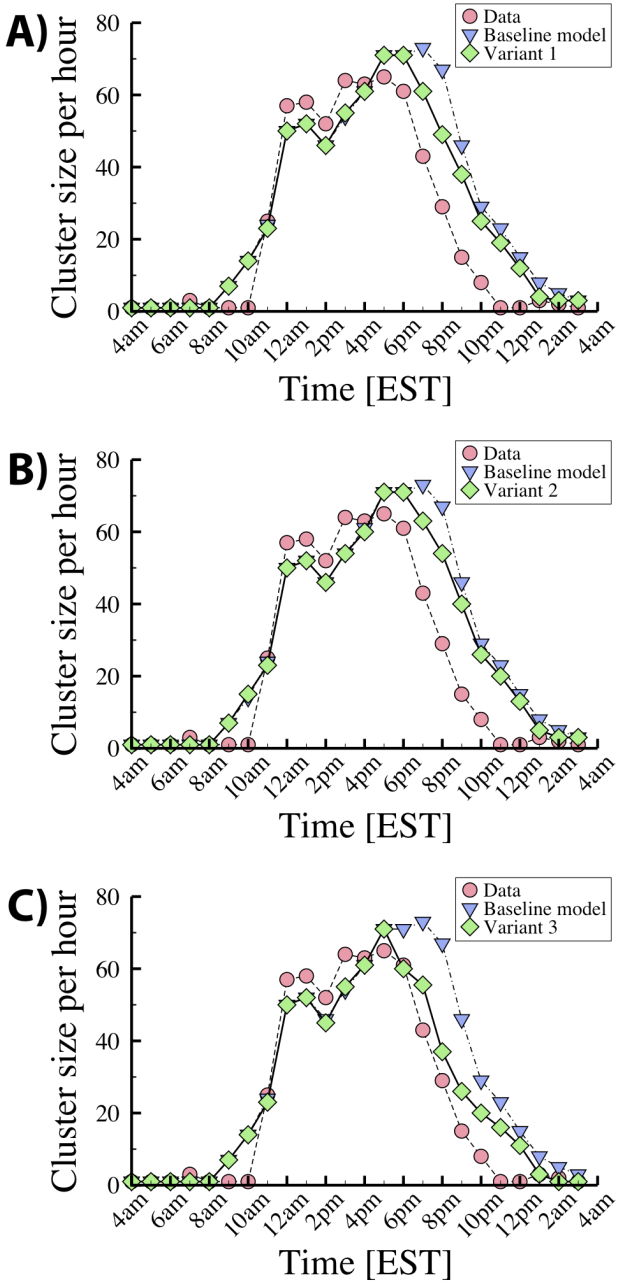

Figure 3. Evolution of the largest cluster size per hour. Comparison between model variants 1 to 3.

Even though the baseline model already shows acceptable results, we intend to improve even further the results by exploring different scenarios. We call these scenarios variants of the model and will go from 1 to 4 (see Table I for a summary of the details). We are interested in improving specially the plateau and the timing of the declining phase of the curves representing the cluster evolution along the day. So far we have introduced the weather disturbances into the baseline model by temporarily reducing the capacity of a few major airports, the next step is to simulate the human interventions on the schedule. Naturally, one way of tackling the congestion problem is flight cancellation. According to our modeling approach, this external intervention should affect the airport network connectivity. The passengers and crew of the canceled flight will not be able to connect with following flights in their destination airports. One way of translating this effect into our model is to temporarily modify the α parameter. Fig. 3-A shows the cluster size evolution for what we refer to as variant 1 of the model and compares the results with real observations and the baseline model. In particular, variant 1 of the model incorporated a connectivity parameter α set to 0 between 7pm and 9pm EST ($\Delta T^\alpha$). This time window (8pm ± 1 hour) is selected because it corresponds to the time with the maximum number of flight cancellations (see Fig. 1-B). As can be seen in Fig. 3-A, the simulation results improve for the declining phase, verifying our assumption of modeling the intervention through a decrease of the network connectivity. We also check the sensitivity of the simulation output to a change in the connectivity factor α and the time window $\Delta T^\alpha$, variant 2 and variant 3 of the model, respectively. In Fig. 3-B, we consider for variant 2 a decrease of α in one half, instead of setting it to zero, and this slightly increases the congestion on the declining phase if compared to variant 1. The effect of an increase in $\Delta T^\alpha$ is even more significant as shown in Fig. 3-C for variant 3 of the model. In this case we fix α to 0 (as in variant 1) increasing the time window by two hours between 6pm and 10pm. Therefore, we can observe a refinement of the declining phase matching. It is important to note that after the $\Delta T^\alpha$ period the cluster size slightly grows and this effect is not seen in the empirical data. This could be due to the fact that queue congestion at the airports does not ease off by this intervention, triggering the propagation of delays when the connectivity is reestablished. The above results show how with slight changes one can gradually improve the capacity of the model to forecast congestion.

After exploring the effect of changing the connectivity on the model, we consider now the implementation of a further response element: the so-called Ground Stops. This consists in preventing the departure of flights on origin when the destination airport has problems [38]. In our case, this measure affects the flights with destination in one of the four airports with reduced capacity and with scheduled arrival time between 8am and 10am Eastern Time (same period of time when β is set to 0 in the baseline mode). The connectivity factor and the airport capacity remain constant throughout the day and equal to the values of the basic model (respectively 0.26 and 1). This scenario corresponds to the variant 4 of the baseline model (Table I). The results of the simulation for the evolution of the largest congested cluster size can be seen in Figure 4. The congestion starts earlier because the delays surge before in time and the largest clusters extension considerably increases. The effect of an early onset of Ground Stops is shown to be devastating for the delay dynamics. Still, generalized Ground Stops is not likely to happen without early palliative interventions such as flight cancellations or diversions.

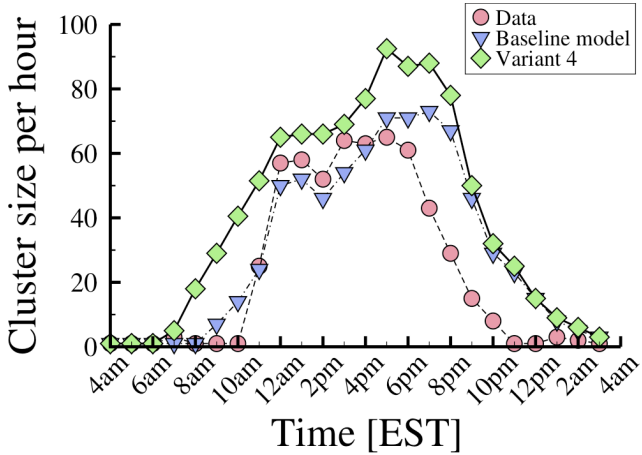

Figure 4.  Evolution of the largest cluster size per hour. Comparison between variant 4 (Ground Stops) and reality.

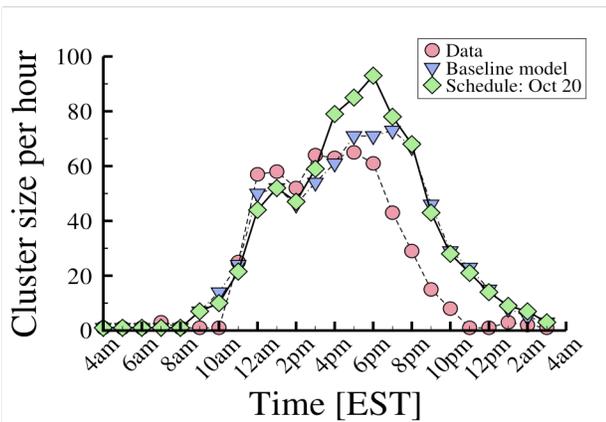

Figure 5.  Evolution of the largest cluster size per hour. Green dots, using the schedule of October 20 with the initial condition of October 27.

In our simulations, we have used the schedule of October 27 as observed in the real operations including flight diversions and cancellations. One may thus wonder which is the effect of this particular configuration of the schedule on the final spreading of the delays and how much the changes in the schedule help to reduce congestion. We have not access to the unperturbed day plan of the airlines but we can still use for the sake of comparison the schedule of October 20. The system this day showed a low level of congestion and so the interventions in the schedule must have been minimal. The variant 4 of the model is thus the baseline scenario, plus the initial conditions of October 27 but implemented in the schedule of October 20. The results of the simulations are depicted in Figure 5. It is clear that the schedule of the 27th was not the reason for the unfolding of a large congestion since an important congested cluster of airports still appears. We can thus blame the weather disturbances for most of the congestion of October 27. The differences between the evolution of the unperturbed schedule of October 20 and the baseline model speak in favor of the intervention measures taken on October 27.

## VI.  CONCLUSIONS

In summary, we analyze the effects that external disturbances and interventions produce in the US air traffic network. In particular, our analysis focuses on October 27 2010 because a large meteorological disruption occurred that day. We count with a data-driven model for the delay propagation in the US network and we implement in the model the schedule information, primary delays and different intervention measures that took place that day. We present the results as a function of a new metric able to measure the level of network-wide extensions of the delays, namely the clusters of congested airports. By computing the evolution of the largest congested cluster size of the day, we compare the empirical results with the delay dynamics observed in the model and find good agreement when weather impacts and canceled flights are considered as input variables. The modeling approach is data-driven in the sense that is based on real records obtained from the US performance data, and agent-based at the level of aircrafts. We introduce the weather impacts by varying the airport capacity parameter $\beta$ to some airports in the network. This change produces a drop in the airport capacity service rate enlarging the airport queue. On the other hand, we implement flight cancellations by affecting the network connectivity parameter $\alpha$, thus reducing the delay propagation dynamics. Our simulations evidence that weather impacts could produce system congestion independently of the day considered, as it is the case when the initial conditions and same input perturbations are introduce to the schedule of October 20.

The methodology employed here is simple but generates, nonetheless, results rich in details that can be used as a predictive tool. Furthermore, our model offers the possibility of evaluating different policy decisions before their real implementation. We show a way of introducing different external inputs that can be used at the strategic planning level to assess possible delay management tools for airports, airlines and the whole network. There are, of course, many possible interventions whose efficiency could be assessed. To give an example, we can quantify the sensitivity of airports to delays, or which ones are most prone to magnify delays, and design palliative measures customized for each airport concentrated, for instance, in increasing the slack time in turnarounds.

## VII.  ACKNOWLEDGMENTS

PF receives support from the network Complex World within the WPE of SESAR (Eurocontrol and EU Commission). JJR acknowledges funding from the Ramón y Cajal program of the Spanish Ministry of Economy (MINECO). Partial support from MINECO and FEDER was received through projects MODASS (FIS2011-24785), FISICOS (FIS2007-60327) and INTENSE@COSYP (FIS2012-30634). Funding was also received from the EU Commission through projects EUNOIA (FP7-DG.Connect-318367) and LASAGNE (FP7-ICT-318132).